\documentclass[%
 amsmath,amssymb,superscriptaddress,reprint,
prb
]{revtex4-1}

\usepackage{graphicx,subfigure}
\usepackage{dcolumn}
\usepackage{bm}
\usepackage{amsthm}

\usepackage{nicefrac}
\usepackage{etex}                  
\usepackage{pgfplots}
\usepackage{tikz}
\usepackage{tikz-3dplot}
\usepgfplotslibrary{fillbetween}      
\pgfplotsset{compat=newest}
\pgfplotsset{plot coordinates/math parser=false}
\usetikzlibrary{calc,fadings,shapes.misc,3d,arrows,
 decorations,decorations.pathmorphing,
 decorations.text,shapes.geometric,intersections,
 decorations.markings,patterns,spy,mindmap}

\preprint{APS/123-QED}

\begin{document}

\title{The energy point of view in plasmonics}

\author{Rabih Ajib}
\affiliation{Universit\'e Clermont Auvergne, CNRS, SIGMA Clermont, Institut Pascal, BP 10448, F-63000 CLERMONT-FERRAND, FRANCE}
\affiliation{Universit\'e Clermont Auvergne, CNRS, Laboratoire de Physique de Clermont, F-63000 CLERMONT-FERRAND, FRANCE}

\author{Armel Pitelet}
\author{R\'emi Poll\`es}
\author{Emmanuel Centeno}
\affiliation{Universit\'e Clermont Auvergne, CNRS, SIGMA Clermont, Institut Pascal, BP 10448, F-63000 CLERMONT-FERRAND, FRANCE}

\author{Ziad Ajaltouni}
\affiliation{Universit\'e Clermont Auvergne, CNRS, Laboratoire de Physique de Clermont, F-63000 CLERMONT-FERRAND, FRANCE}

\author{Antoine Moreau}%
\affiliation{Universit\'e Clermont Auvergne, CNRS, SIGMA Clermont, Institut Pascal, BP 10448, F-63000 CLERMONT-FERRAND, FRANCE}
 \email{antoine.moreau@uca.fr}

\begin{abstract}
The group velocity of a plasmonic guided mode can be written as the ratio of the flux of the Poynting to the integral of the energy density along the profile of the mode. This theorem, linking the way energy propagates in metals to the properties of guided modes and Bloch modes in a multilayer, provides a unique physical insight in plasmonics. It allows to better understand the link between the negative permittivity of metals and the wide diversity of exotic phenomenon that occur in plasmonics -- like the slowing down of guided modes, the high wavevector and the negative refraction. 
\end{abstract}

\maketitle

Deeply subwavelength metallic structures give us an unprecedented control of visible light, allowing to focus, concentrate, absorb, scatter light very efficiently, or to even enhance the emission of light by fluorophores\cite{zhukovsky2013physical} at totally new levels. Metals actually present a very peculiar optical response that dielectrics are totally incapable of - which can be linked to the presence of a free electron gas, a plasma, inside even the tiniest metallic nanoparticles\cite{drude1900elektronentheorie}.

Plasmonic resonators are the smallest optical resonators possible and their resonances can always be linked to the excitation of some kind of plasmonic guided mode. There is thus a large diversity of these modes, from the well known surface plasmon\cite{homola2008surface} to long and short-range surface plasmons\cite{berini2009long}, gap-plasmons\cite{boz} or modes supported by hyperbolic metallo-dielectric multilayers\cite{PhysRevB.75.241402}. Most of them present very high wavevectors which explains the reduced size of the plasmonic resonators\cite{moreau12b,smaali2017miniaturizing}: they are essentially cavities for guided modes with very small effective wavelength. One must finally underline that exotic phenomena like negative refraction may also occur in metallo-dielectric multilayers\cite{poddubny2013hyperbolic,benedicto11b,xu2013all,centeno2015effective}.

All these features lack a unified view that would enable to give a physical insight into the reasons why large wavevector guided modes and negative refraction are common in plasmonics and very exotic in dielectric structures - requiring the careful tailoring of photonic crystals, for instance\cite{krauss2007slow,lalanne2018structural}. We think that considering the way that energy flows when such modes propagate provides this kind of insight.

The average flux of the Poynting vector has actually been used in the context of metamaterial and negative index materials as a useful tool to predict in which direction a mode really propagates ({\em i.e.} the sign of its group velocity). Such a approach relies on a theorem showing that the energy velocity is equal to the group velocity for modes propagating in non-dispersive, dielectric media\cite{yeh2005optical,yariv1977electromagnetic}. This theorem has been mostly ignored because, except in a few cases like when a mode approaches a cut-off condition, the group velocity does not present any exotic behaviour.

Assuming this link holds even in the case of plasmonic or metamaterial waveguides, it can prove very useful to determine the sign of the group velocity by simply using the profile of the guided modes at a given frequency without having to actually compute the dispersion relation\cite{tournois1997negative,polles10,benedicto11a}. However, the optical response of metals is linked to the presence of free electrons, that transport a part of the guided mode energy and whose kinetic energy can not be neglected. For a plane wave propagating in a plasma, provided the energy of the electrons is taken into account both in the energy flux and in the energy density, it has been shown by Bers\cite{bers2000note} that the energy and flux velocity are the same. This underlines that Yariv and Yeh's theorem\cite{yeh2005optical} can not be applied to plasmonic waveguides. Since metals are highly dispersive and can be considered as boxes containing a real plasma, it is even surprising that computing the average flux of the Poynting vector could be successful in predicting the sign of the group velocity.

Here we show that it is actually possible to generalize Yariv and Yeh's theorem in the context of dispersive media, including metals.
This means that there is no need to modify the expression of the energy flux and only to adapt the energy density expression, to make the theorem valid -- despite what has been established for plasma\cite{bers2000note}. Said otherwise, the group velocity of a guided mode in a plasmonic multilayer is equal to the energy velocity of the electromagnetic field alone, and the energy transported by the free electrons, while not negligible\cite{bers2000note}, can be completely ignored. We then show using several examples how considering the energy velocity can provide a unifying vision of the optical response of plasmonic multilayers.

We consider a multilayered structure invariant in the $x$ and $y$ directions, and a guided mode, solution of Maxwell's equations presenting a $e^{i(k_x\,x-\omega\,t)}$ dependency in $x$ and $t$. 
We will assume the mode is p-polarized, because nothing exotic occurs for the s polarization in metallo-dielectric structures. Maxwell's equations reduce to 

\begin{align}
\partial_z E_x - ik_x E_z =& i\omega \mu_0 H_y\label{max1}\\
\partial_z H_y =&i\omega \epsilon_0\epsilon E_x\label{max2}\\
ik_x H_y=&-i\omega \epsilon_0 \epsilon E_z\label{max3}
\end{align}

\newcommand{\V}[1]{{\bf #1}}
\newcommand{\de}{\delta}

Any change in the mode will be linked to a small change in its propagation constant, noted $\delta k_x$, its pulsation $\delta \omega$, its electric and magnetic field, respectively $\delta\V{E}$ and $\delta \V{H}$. These small changes are all linked by Maxwell's equations, whatever the dispersion relation of the guided mode which is considered. These equations can thus be differentiated to yield 
\begin{align}
-i\,\delta k_x\,E_z - ik_x \,\delta E_z + \partial_z\,\delta E_x=&i\,\delta\omega\,\mu_0 H_y + i\omega\mu_0\,\delta H_y\label{dmax1}\\
\partial_z \,\delta H_y =& i\,\delta\omega \,\epsilon_0 \epsilon \,E_x + i\omega \epsilon_0 \,\delta \epsilon\,E_x \nonumber \\&+ i\omega\epsilon_0\epsilon\delta E_x\label{dmax2}\\
i\,\delta k_x\,H_y + ik_x\,\delta H_y =&-i\,\delta\omega\,\epsilon_0\epsilon E_z-i\omega\epsilon_0\,\delta\epsilon\, E_z \nonumber\\&- i \omega \epsilon_0\epsilon\,\delta E_z\label{dmax3}
\end{align}
and since $\epsilon$ is only a function of $\omega$, we can write that $\de \epsilon = \de \omega \,\frac{\partial_\epsilon}{\partial \omega}$.

Following Yariv and Yeh\cite{yeh2005optical,yariv1977electromagnetic}, we introduce now the quantity
\begin{equation}
\V{F} = \delta \V{E}\otimes \V{H}^*+\delta \V{H}^*\otimes\V{E}+\V{H}\otimes \delta \V{E}^*+\V{E}^*\otimes \delta\V{H},
\end{equation}
where ${}^*$ denotes the complex conjugate. 

Since we restrain ourselves here to a multilayered structure, we only need to calculate $\partial_z\, F_z = 2i \partial_z \Im \left( \delta E_x\,H_y^* - E_x\,\delta H_y^*\right)$. Given its expression, we calculate the quantity
\begin{align*}
\mathbb{A}=&\partial_z \left(\delta E_x\,H_y^* - E_x\,\delta H_y^*\right)\\
=& \partial_z \de E_z . H_y^* + \de E_x . \partial_z H_y^* -\partial_z \,\de H_y^* . E_x + \de H_y^* . \partial_z E_x\\
\end{align*}
Using respectively \eqref{dmax1},\eqref{dmax2} and \eqref{dmax3} we find that the different terms can be written
\begin{align}
\partial_z(\de E_x) H_y^* =& i\,\de k_x\,E_zH_y^* + ik_x H_y^*\,\de E_z + i\,\de \omega\,\mu_0 |H_y|^2 \nonumber\\&+ i\omega\mu_0\,\de H_y\,H_y^*\\
\partial_z H_y^* . \de E_x =& -i\omega\epsilon\epsilon E_x^*\,\de E_x\\
-\partial_z(\de H_y^*) E_x =& i\,\de\omega\,\epsilon_0\left(\epsilon + \omega\,\frac{\partial \epsilon}{\partial \omega}\right) |E_x|^2 + i\omega \epsilon_0\epsilon\,\de E_x^*\,E_x\\
-\de H_y^*\,\partial_z E_x=&-ik_xE_z\,\de H_y^* + i\omega \mu_0 H_y\,\de H_y^*
\end{align}

Using \eqref{max3} and \eqref{dmax3}, we have in addition 
\begin{align}
ik_x \left(\de E_z H_y^* - E_z\, \de H_y^*\right) =& \de E_z \,i\omega\epsilon_0\epsilon E_z^* + i\omega\epsilon_0\epsilon\,\de E_z^*\,E_z\nonumber\\&+ \de k_x\, E_z H_y^* + i\omega\epsilon_0\left(\epsilon +\omega\,\frac{\partial \epsilon}{\partial \omega}\right) |E_z|^2. \label{large}
\end{align}

Adding all the terms to calculate $A$ and using \eqref{large}, we finally get 
\begin{align}
\partial_z F_z &= 4i\,\de k_x \Re \left(E_z H_y^*\right)  \nonumber\\& + 2i\,\de \omega \left( \mu_0  |\V{H}|^2+\epsilon_0 \left\{ \epsilon +\omega\,\frac{\partial \epsilon}{\partial \omega} \right\} |\V{E}|^2\right)
\end{align}
where all the real terms have been eliminated.

Now if we consider a mode guided along the $x$ axis in a multilayered structure containing metallic layers, then the radiation condition impose vanishing fields amplitudes at infinity and thus a vanishing $F_z$, which allows to write that over a section of the waveguide we have 
\begin{align}
\int_{-\infty}^{+\infty} \partial_z F_z\, \mbox{d}z =0.
\end{align}
This allows to conclude that the group velocity $v_g$ is given by
\begin{align}
v_g = \frac{\delta \omega}{\delta k_x} = \frac{-\int \frac{1}{2}\Re \left(E_z H_y^*\right)\,\mbox{d}z}{\int \frac{1}{4}  \mu_0  |\V{H}|^2+\frac{1}{4}\epsilon_0 \left\{ \epsilon +\omega\,\frac{\partial \epsilon}{\partial \omega} \right\} |\V{E}|^2 \mbox {d}z}\label{eq:ve}
\end{align}
that is, the ratio of the total x-directed time averaged Poynting vector, in the numerator, to the total time averaged energy, in the denominator. 

We underline that such a proof naturally yeld the classical expression of the energy density in a dispersive media whose permittivity depends on the frequency -- so that this constitutes a fourth way, after the approaches of Brillouin, Landau and Loudon\cite{nunes2011electromagnetic} to reach this expression. This way may even be the most natural.

\begin{figure}\centering
\includegraphics[]{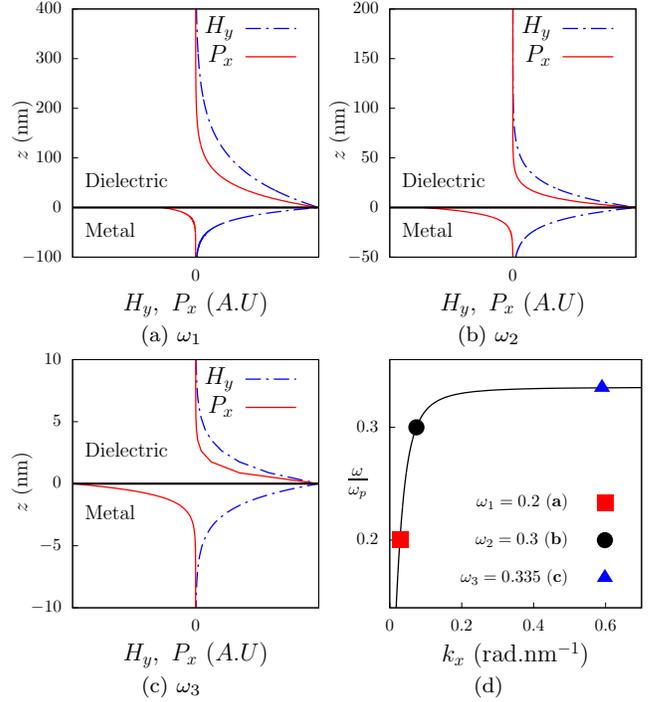}
\caption{\textbf{(a)}, \textbf{(b)}, \textbf{(c)} Magnetic field and Poynting vector profiles in the case of the surface plasmon for different values of $\frac{\omega}{\omega_p}$. \textbf{(d)} Dispersion curves for the surface plasmon, showing the frequencies $\omega_1$, $\omega_2$ and $\omega_3$.\label{fig:sp}}
\end{figure}

\begin{figure}\centering
\includegraphics[]{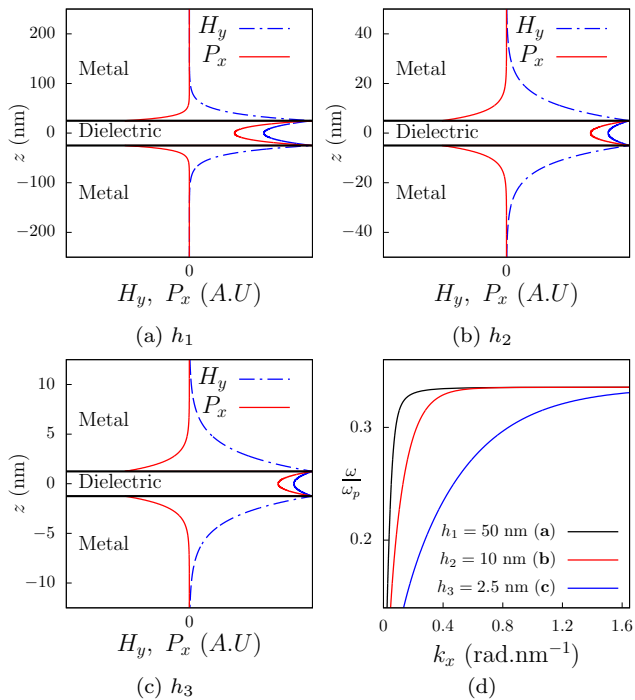}
\caption{\textbf{(a)}, \textbf{(b)}, \textbf{(c)} Magnetic field and Poynting vector profiles of the gap plasmon for different values of the dielectric gap width $h$ and $\frac{\omega}{\omega_p}=0.25$. \textbf{(d)} Dispersion curves in the case of the gap plasmon, for, from left to right, $h_1$, $h_2$ and $h_3$.\label{fig:gp}}
\end{figure}

In order to illustrate when the theorem provides a better understanding of plasmonics in general, we first consider the emblematic surface plasmon propagating at the interface between a metal with a permittivity $\epsilon_m$ and a dielectric with a permittivity $\epsilon_d$. Surface plasmon dispersion relation reads 
\begin{align}
 \frac{\kappa_m}{\epsilon_m}+\frac{\kappa_d}{\epsilon_d} = 0,
\end{align}
where $\kappa_i=\sqrt{(k_x^2-k_0^2\epsilon_i)}$ are wavevectors in the $\hat{z}$ direction defined such that the waves are 'propagating' away from interface as $e^{\pm\kappa_z z}$. To compute $v_g$ one only need the mode wavevector $k_x$ by solving dispersion relation and from it, the knowledge of fields amplitudes implied in \eqref{eq:ve} allow interpretation of $v_g$ in term of energy. In this way we have numerically checked the validity of \eqref{eq:ve} ensuring $v_g=v_E$. The mode profile, as well as the Poynting vector and the dispersion curve are shown on Fig. \ref{fig:sp}. Since the Poynting vector is proportional to $\frac{1}{\epsilon} |H_y|^2$ and since $\epsilon_m<0$, the energy flux is negative in the metal and positive in the dielectric. Far below the plasma frequency, this phenomenon has no real impact on the propagation: the permittivity is very large, the negative flux is thus very small. When the frequency gets closer to the plasma frequency, the permittivity in the metal decreases, the negative flux of the energy increases and the energy velocity of the whole mode is thus decreasing. When the frequency is approaching $\omega_{sp} = \frac{\omega_p}{\sqrt{1+\epsilon_d}}$ the permittivity in the metal is negative but close to the permittivity of the dielectric in absolute value. Since the magnetic field is continuous at the interface, this means that the negative flux almost balances the positive one. The energy velocity thus goes to zero, and thanks to the theorem above, we know the group velocity does too. Since the group velocity is the inverse of $\frac{\partial k_x}{\partial \omega}$ this means that this quantity is increasing and finally $k_x$ is thus increasing when the frequency approaches $\omega_{sp}$. There is thus a direct link between the fact that the Poynting flux vanishes and the high wavevector presented by the surface plasmon. This effect in extreme in the sense that the phase velocity vanishes too, and the mode approaches cut-off condition. More generally, as light propagates close to a metal, the energy propagates backward in the metal, which slows down the propagation of light itself. Light thus experiences what we call a plasmonic drag.

This phenomenon is even more obvious in the case of the gap-plasmon\cite{boz}, because a geometrical parameter (instead of the frequency) allows to control the Poynting balance. A gap-plasmon is a mode in a dielectric sandwiched between two metals. The dispersion relation of this guided mode reads

\begin{align}
 \frac{\kappa_m}{\epsilon_m}+\frac{\kappa_d}{\epsilon_d}\tanh (\frac{\kappa_d h}{2}) = 0.
\end{align}

The thickness of the dielectric $h$ is small enough so that the picture of coupled surface plasmons does not hold any more\cite{moreau13}. The fundamental mode presents a diverging wavevector when the gap goes to zero. Fig. \ref{fig:gp} shows the profile of a gap-plasmon, its Poynting vector and dispersion curves for different gap width. The whole behavior of the mode is easier to understand from the energy point of view: when the gap closes or when the frequency is getting close to $\omega_{sp}$ the energy flux in the metal begins to balance the flux in the dielectric. The mode is thus slowed down. The dispersion curve of the gap-plasmon resemble to the one of a surface plasmon, but when the gap closes the energy (and thus the group) velocity goes to zero, pushing the wavevector to correspondingly larger values.

The same reasoning can be applied to more complex structures, like the multilayers alternating dielectric layers of thickness $h_d$ and metallic ones with a thickness $h_m$. The plasmonic drag is largely present too.  The modes in that case present very high wavevector when the ratio $\frac{h_m}{h_d}$ is large enough and have the advantage of propagating in a thicker structure compared to the gap-plasmon. They are thus easier to excite using end-fire coupling methods, and the resonators that can be obtained using such structures (hyperbolic wire antennas) are deeply subwavelength while preserving their cross-section\cite{smaali2017miniaturizing}. Dispersion relation and amplitudes of guided modes in such structures, or in any arbitrary multilayer can be found by solving an eigenvalue problem based on the transfer\cite{dacheng16} or scattering\cite{defrance16} matrix methods then allowing an energy point of view interpretation thanks to \eqref{eq:ve}. An homogenization procedure even leads to simplified expressions for the wavevector and the field amplitudes\cite{smaali2017miniaturizing} for such modes.

Finally, we would like to underline that, as in the original work of Yariv and Yeh, the theorem can be applied not only to guided modes, but to Bloch modes too. For Bloch modes in periodical structures\cite{yariv1977electromagnetic}, the same conclusion can be reached except that the integration has only to be done on one period of the multilayer, whatever its complexity\cite{xu2013all}. We consider now a multilayer with a period composed of a metallic layer with a thickness $h_m$ and a dielectric layer with a thickness $h_d$.  Using the periodicity $D=h_m+h_d$, the theorem can be written 
\begin{align}
v_g = \frac{-\frac{1}{D}\int_{0}^{D} \frac{1}{2}\Re \left(E_z H_y^*\right)\,\mbox{d}z}{\frac{1}{D}\int_{0}^{D} \frac{1}{4}  \mu_0  |\V{H}|^2+\frac{1}{4}\epsilon_0 \left\{ \epsilon +\omega\,\frac{\partial \epsilon}{\partial \omega} \right\} |\V{E}|^2 \mbox {d}z},\label{eq:veBloch}
\end{align}

In that case, the results would be very similar to what has been published recently in the case of periodical lossy structures\cite{wolff2018modal}. This allows to better understand when such a mode will present a negative group velocity for instance. Such a phenomenon emerges when the energy and the group velocity are opposite to the wavevector along the interfaces. In the limit where the layers are all very thin compared to the wavelength, the homogenization regime, the magnetic field does not really change from one layer to the other, given how thin they are. In that case, the numerator of \eqref{eq:veBloch} can be recast as 
\begin{align}
-\frac{1}{D}\int_{0}^{D} \frac{1}{2}\Re \left(E_z H_y^*\right)\,\mbox{d}z & =\frac{1}{D}\frac{k_x}{2\omega \epsilon_0}\left(\frac{h_m}{\epsilon_m}+\frac{h_d}{\epsilon_d} \right)|H_y|^2, \\
& = \frac{k_x}{2\omega \epsilon_0\epsilon_{eff} } |H_y|^2,
\end{align}
with 
 $\epsilon_{eff}=\frac{D}{\frac{h_m}{\epsilon_m}+\frac{h_d}{\epsilon_d}}$
which is precisely the $zz$ component of an effective permittivity tensor corresponding to an equivalent homogeneous anisotropic medium for the periodic multilayer\cite{zhukovsky2013physical}. 

When the above quantity is negative then group velocity and the wavevector along the $x$ axis signs are opposite, leading to negative refraction. That is, energy along the interfaces propagate in the opposite direction of the impinging wavevector $x$ component - the dispersion curve is in that case hyperbolic\cite{centeno2015effective}. The condition $\epsilon_{eff}<0$ can be written
\begin{equation}
\frac{h_m}{\epsilon_m}+\frac{h_d}{\epsilon_d}<0
\end{equation}
and under this form, it can be interpreted as a simple Poynting balance over one period: the Poynting flux in the metal is proportional to $\frac{h_m}{\epsilon_m}$ while the Poynting flux in the dielectric layer is proportional to $\frac{h_d}{\epsilon_d}$ because $H_y$ is essentially constant. The condition for which negative refraction occurs can thus be seen as equivalent to requiring the global Poynting flux to be opposite to the wavevector. This example shows how general the vision of plasmonics through the prism of energy can be. We underline that this also allows to easily understand why dielectric multilayers are completely unable to produce negative refraction: there is no way the average Poynting flux can be negative when all the permittivities are positive\cite{benedicto11a}, which underlines how peculiar the response of plasmonic multilayers is in comparison.

In conclusion, we have extended Yariv and Yeh's theorem to dispersive media, allowing the expression for the energy density in dispersive media to appear naturally, and shown its importance in the framework of plasmonics by illustrating it on various examples. The theorem shows that even in plasmonics where electrons store a lot of the energy, the energy velocity of the electromagnetic wave alone is equal to the group velocity. This probably means that there should be a way to define an energy flux and an energy velocity that take into account the free electrons, and probably to get an equivalent result\cite{bers2000note}, but this is beyond the scope of the present paper -- and it would not be as a powerful tool to understand plasmonics. Considering the way the energy flows, through the calculation of the average Poynting flux essentially, actually provides a physical picture that spans the whole zoology of plasmonic guided modes. The Poynting vector along the propagation direction is indeed negative in metals, leading to a slowing down of any light propagating in their vicinity. When this plasmonic drag is extreme, it leads to very small group velocity and large wavevectors, a crucial parameter to explain the extraordinary way metals can concentrate light in deeply subwavelength volumes. When the energy flux in metals overwhelms the one in the dielectric, as has been shown in metallo-dielectric structures, negative refraction occurs. While we don't expect this vision to allow any new discovery in such a thoroughly studied field, we think it really explains why the properties of metals in the plasmonic regime, characterized by relatively low absolute values of the negative permittivity of metals, are so peculiar.

\bibliography{sample}
\end{document}